\documentclass{article}
\usepackage{amsmath}

\numberwithin{equation}{section}
\input{tcilatex}
\begin{document}

\centerline{\large\bf Stochastic local operations and classical
communication } 
\centerline{\large\bf  equations and
classification of even $n$ qubits\footnote{The paper was supported by
NSFC(Grants No.
 10875061,60433050, and 60673034 ) and Tsinghua National Laboratory for
Information Science and Technology. }} 

\vspace*{8pt}
\centerline{Xiangrong Li$^{a}$, Dafa Li$^{b}$}
\vspace*{5pt}

\centerline{$^a$ Department of Mathematics, University of
California, Irvine, CA 92697-3875, USA}

\centerline{$^b$ Department of mathematical sciences, Tsinghua
University, Beijing 100084 CHINA}


\abstract
For any even $n$ qubits we establish four SLOCC equations and
construct four SLOCC polynomials (not complete) of degree $2^{n/2}$, which
can be exploited for SLOCC classification (not complete) of any even $n$
qubits. In light of the SLOCC equations, we propose several different
genuine entangled states of even $n$ qubits and show that they are
inequivalent to the $|GHZ\rangle $, $|W\rangle $, or $|l,n\rangle $ (the
symmetric Dicke states with $l$ excitations) under SLOCC via the vanishing
or not of the polynomials. The absolute values of the polynomials can be
considered as entanglement measures.



\section{Introduction}

A fundamental concept in quantum information theory is the understanding of
entanglement. Quantum entanglement can be viewed as a crucial resource in
quantum information. The key question is how to quantify and classify
entanglement of quantum states. Polynomial functions in the coefficients of
pure states which are invariant under stochastic local operations and
classical communication (SLOCC) transformations have been studied extensively 
\cite{Wootters,Coffman,Miyake,Sharma,Luque1,Luque2,Djokovic,Wong,LDF10,LDFntangle,LDF07a,LDFJMP09,Osterloh1,Osterloh2,Leifer,Levay}
and exploited to construct entanglement measures 
\cite{Wootters,Coffman,Luque1,LDFJMP09,Osterloh1,Osterloh2}. The concurrence 
\cite{Wootters} and three-tangle \cite{Coffman}, which measure entanglement 
of two-qubit and three-qubit states, are polynomial invariants of degrees 2 
and 4 respectively. It is known that the concurrence and three-tangle are
the absolute values of hyperdeterminants for two and three qubits
respectively \cite{Miyake}. 
An expression has recently been derived for four-tangle, which is
a polynomial invariant and a measure of genuine entanglement of four-qubit
states \cite{Sharma}.
Polynomial invariants of degrees 2, 4 and 6 for four and
five qubits have been constructed from classical invariant theory \cite%
{Luque1,Luque2}. The absolute values of the polynomial invariants obtained
in \cite{Luque1} may be used to construct entanglement measures of
four-qubit states. Further, polynomial invariants of degrees 2, 4, 6, 8, 10
and 12 for four and five qubits have been obtained using local invariant
operators \cite{Djokovic}. Despite these efforts, few attempts have so far
been made towards the generalization to higher number of qubits.
Three-tangle has been generalized to $n$-tangle for even $n$ qubits \cite%
{Wong} and has been shown to be equal to the square of the polynomial
invariant of degree 2 \cite{LDF10}. A generalization of three-tangle to odd $%
n$ qubits has been recently proposed in \cite{LDFntangle}. In \cite{LDF07a},
polynomial invariants of degree 2 for even $n$ qubits and degree 4 for odd $n
$ qubits  have been derived by induction based on the definition of SLOCC.

SLOCC classification of pure states has been under intensive research \cite%
{Miyake,Dur,Verstraete,Lamata07,Borsten,Chterental,Cao,LDF07b,LDFQIC09,Buniy,Viehmann,Bastin,LDFEPL09}%
. For three qubits, two genuine entanglement states, namely the $|GHZ\rangle
$ and $|W\rangle$ states, have been distinguished and characterized by the
vanishing or not of the three-tangle \cite{Dur}. For four or more qubits,
the number of SLOCC classes is infinite. It is highly desirable to divide
these infinite SLOCC classes into a finite number of families. Central to
the issue is the criteria to determine which family an arbitrary state
belongs to. Various methods have been undertaken to tackle the
classification of four-qubit states, including those based 
on Lie group theory \cite{Verstraete}, on hyperdeterminant \cite{Miyake}, on inductive approach \cite{Lamata07}, on string theory \cite{Borsten}, 
and on polynomials (algebraic) invariants 
\cite{Chterental,Cao,LDF07b,LDFQIC09,Buniy,Viehmann}. Recently, the Majorana
representation has been used for SLOCC entanglement classification of $n$%
-qubit symmetric states \cite{Bastin}. For $n$ qubits, it is known that the $%
|l,n\rangle $ states (symmetric Dicke states with $l$ excitations) are
inequivalent to the $|GHZ\rangle$ state or the $|W\rangle$ state under SLOCC 
\cite{LDFEPL09}. Therefore it is necessary to develop schemes to find other
genuine entangled states which are inequivalent to the $|GHZ\rangle $, $%
|W\rangle $, or $|l,n\rangle $ states.

In this paper, we establish four SLOCC equations and construct four SLOCC
polynomials (not complete) of degree $2^{n/2}$ for any even $n$ qubits. The
equations are obtained from the polynomials (determinants) of the
coefficients of the two SLOCC equivalent states by induction via direct
manipulation of SLOCC definition. For $n=4$, the SLOCC polynomials of degree 
$2^{n/2}$ reduce to the polynomials of degree 4 in \cite{Luque1}. In light
of the SLOCC equations, we propose several different genuine SLOCC
entanglement classes of even $n$ qubits and show that they are inequivalent
to the $|GHZ\rangle $, $|W\rangle $, or $|l,n\rangle $ (the symmetric Dicke
states with $l$ excitations) SLOCC classes via the vanishing or not of the
polynomials.

The manuscript is organized as follows. In Sections 2, 3, 4 and 5, we
construct SLOCC polynomials and present SLOCC equations of type I, II, III,
and IV, respectively. We also discuss SLOCC classifications by means of
these polynomials. In Section 6, we draw our conclusions. 

\section{SLOCC equation and polynomial of type I}

Let $|\psi \rangle $ and $|\psi ^{\prime }\rangle $ be any states of $n$
qubits. Then we can write 
\begin{equation*}
|\psi ^{\prime }\rangle =\sum_{i=0}^{2^{n}-1}a_{i}|i\rangle ,|\psi \rangle
=\sum_{i=0}^{2^{n}-1}b_{i}|i\rangle,
\end{equation*}
where $\sum_{i=0}^{2^{n}-1}|a_{i}|^{2}=1$ and $%
\sum_{i=0}^{2^{n}-1}|b_{i}|^{2}=1$. Two states $|\psi \rangle $ and $|\psi
^{\prime }\rangle $ are equivalent under SLOCC if and only if there exist
invertible local operators $\mathcal{A}_{1}$, $\mathcal{A}_{2}, \cdots , 
\mathcal{A}_{n}$ such that 
\begin{equation}
|\psi ^{\prime }\rangle =\underbrace{\mathcal{A}_{1}\otimes \mathcal{A}%
_{2}\otimes \cdots \otimes \mathcal{A}_{n}}_{n}|\psi \rangle.
\label{neqcond}
\end{equation}

For the state $|\psi ^{\prime }\rangle $ of even $n$ qubits, let $\Theta
(a,n)$ be the determinant of the coefficient matrix ($2^{n/2}$ by $2^{n/2})$
which is partitioned into blocks, \emph{i.e.} 
\begin{equation}
\Theta (a,n)=\left\vert \left( 
\begin{tabular}{llll}
$\Theta _{1}$ & $\Theta _{2}$ & $\cdots$ & $\Theta _{2^{n/2}}$%
\end{tabular}%
\right) \right\vert ,
\end{equation}%
where the blocks $\Theta _{i}$, $i=1, \cdots , 2^{n/2}$, are the columns of
the matrix and $\bigl\{\Theta _{1}^{T}$ $\Theta _{2}^{T} \cdots \Theta
_{n}^{T}\bigr\}$ is just the coefficient vector $\bigl\{a_{0}$ $a_{1} \cdots
a_{2^{n}-1}\bigr\}$.

To understand the structure of $\Theta (a,n)$, we list $\Theta (a,4)$ below: 
\begin{equation}
\Theta (a,4)=\left\vert 
\begin{tabular}{llll}
$a_{0}$ & $a_{4}$ & $a_{8}$ & $a_{12}$ \\ 
$a_{1}$ & $a_{5}$ & $a_{9}$ & $a_{13}$ \\ 
$a_{2}$ & $a_{6}$ & $a_{10}$ & $a_{14}$ \\ 
$a_{3}$ & $a_{7}$ & $a_{11}$ & $a_{15}$%
\end{tabular}%
\right\vert ,
\end{equation}%
which turns out to be the determinant $L$ in \cite{Luque1}.

Now, suppose that $|\psi ^{\prime }\rangle $ and $|\psi \rangle $ are
equivalent under SLOCC. Then we get the following result: 
\begin{equation}
\Theta (a,n)=\Theta (b,n)\bigl[\det (\mathcal{A}_{1})\cdots \det (\mathcal{A}%
_{n})\bigr]^{2^{(n-2)/2}},  \label{main-eq1}
\end{equation}%
where $\Theta (b,n)$ is obtained from $\Theta (a,n)$ by replacing $a$ by $b$%
. Eq. (\ref{main-eq1}) and $\Theta (a,n)$ are referred to as SLOCC equation
and polynomial of type I for even $n$ qubits, respectively. The proof of Eq.
(\ref{main-eq1}) for $n=2$ can be seen as follows. Solving Eq. (\ref{neqcond}%
) yields $a_{0}a_{3}-a_{1}a_{2}=$ $(b_{0}b_{3}-b_{1}b_{2})\det (\mathcal{A}%
_{1})\det (\mathcal{A}_{2})$ \cite{LDF07a}. The desired result then follows
by noting that $(a_{0}a_{3}-a_{1}a_{2})$ is the determinant $\Theta (a,2)$
of the coefficients of states for two qubits. For $n\geq 4$, we refer the
reader to Appendix A for the proof.

It follows from Eq. (\ref{main-eq1}) that if one of $\Theta (a,n)$ and $%
\Theta (b,n)$ vanishes while the other does not, then the state $|\psi
^{\prime }\rangle $ is not equivalent to the state $|\psi \rangle $ under
SLOCC.

We next demonstrate that $\Theta (a,n)$ vanishes for the $|GHZ\rangle $, $%
|W\rangle $ and Dicke states for $n>2$. It is trivial to see that $\Theta
(a,n)$ vanishes for the $|GHZ\rangle $ and $|W\rangle $ states. Recall that
the $n$-qubit symmetric Dicke states with $l$ excitations, where $1\leq
l\leq (n-1)$, were defined as \cite{Stockton} 
\begin{equation}
|l,n\rangle =\sum_{i}P_{i}|1_{1}1_{2}\cdots 1_{l}0_{l+1}\cdots 0_{n}\rangle ,
\label{Dicke}
\end{equation}%
where $\{P_{i}\}$ is the set of all the distinct permutations of the qubits.
Note that $|1,n\rangle $ is just $|W\rangle $. For Dicke states $|l,n\rangle 
$, it is known that $|l,n\rangle $ and $|(n-l),n\rangle $ are equivalent to
each other under SLOCC. Hence we only need to consider $2\leq l\leq n/2$.
Inspection of the binary form of the subscripts of the entries in the second
and third columns of $\Theta (a,n)$ reveals that those two columns are
equal. Indeed, for $l<n/2$, we see that all the entries in the last column
of $\Theta (a,n)$ vanish. It follows that $\Theta (a,n)$ vanishes for Dicke
states as well.

Consider the following two states 
\begin{eqnarray}
|\chi _{1}\rangle &=& (1/\sqrt{2^{n/2}}) \biggl[%
\sum_{m=0}^{2^{n/2}-2}|(2^{n/2}+1)m\rangle -|2^{n}-1\rangle\biggr], \\
|\chi _{2}\rangle &=&(1/\sqrt{2^{n/2}})\biggl[\sum_{m=1}^{2^{n/2}-1}
|(2^{n/2}-1)m\rangle -|2^{n}-2^{n/2}\rangle\biggr].
\end{eqnarray}
We observe that all the non-zero coefficients of $|\chi _{1}\rangle $ lie on
the diagonal of $\Theta (a,n)$. This leads to non-vanishing $\Theta (a,n)$
for $|\chi _{1}\rangle $. Similary, all the non-zero coefficients of $|\chi
_{2}\rangle $ lie on the antidiagonal of $\Theta(a,n)$ and therefore $\
\Theta (a,n)$ does not vanish for $|\chi _{2}\rangle $. In light of Eq. (\ref%
{main-eq1}), for $n>2$, $|\chi _{1}\rangle $ and $|\chi _{2}\rangle $ are
both different from the $|GHZ\rangle $, $|W\rangle $ and Dicke states under
SLOCC. It can be further demonstrated that $|\chi _{1}\rangle $ and $%
|\chi_{2}\rangle $ are entangled, and that $|\chi _{2}\rangle $ is
equivalent to $|\chi _{1}\rangle $ under SLOCC. We exemplify the result for
the case of four qubits. We find that $|\chi _{1}\rangle =(1/2)\bigl(%
|0\rangle +|5\rangle +|10\rangle -|15\rangle\bigr)$ and it was shown in \cite%
{LDF07b} that $|\chi _{1}\rangle $ is different from the $|GHZ\rangle $, $%
|W\rangle $, and Dicke states under SLOCC.

\textsl{Remark 2.1. }In $|\chi _{1}\rangle $ SLOCC entanglement class, the
states $|\chi _{1}\rangle $ and $|\chi _{2}\rangle $ have the minimal number
of product terms (\emph{i.e.} $2^{n/2}$ product terms).

\section{SLOCC equation and polynomial of type II}

For the state $|\psi ^{\prime }\rangle $ of even $n$ qubits, let $\Pi (a,n)$
be the determinant of the coefficient matrix ($2^{n/2}$ by $2^{n/2})$ which
is partitioned into blocks, \emph{i.e.} 
\begin{equation}
\Pi (a,n)=\left\vert \left( 
\begin{tabular}{c}
$\Pi _{1}$ \\ 
$\Pi _{2}$ \\ 
$\vdots $ \\ 
$\Pi _{2^{n/2}}$%
\end{tabular}%
\right) \right\vert ,
\end{equation}%
where the blocks $\Pi _{i}$, $i=1, \cdots , 2^{n/2}$, are the rows of the
matrix, $\bigl\{\Pi _{1}$ $\Pi _{3} \cdots \Pi _{2^{n/2}-1}\bigr\}$ is just
the coefficient vector $\bigl\{a_{0}$ $a_{2} \cdots a_{2k} \cdots a_{2^{n}-2}%
\bigr\}$, and $\bigl\{\Pi _{2}$ $\Pi _{4} \cdots \Pi _{2^{n/2}}\bigr\}$ is
just the coefficient vector $\bigl\{a_{1}$ $a_{3}\cdots a_{2k+1} \cdots
a_{2^{n}-1}\bigr\}$.\ 

To understand the structure of $\Pi (a,n)$, we list $\Pi (a,4)$ below: 
\begin{equation}
\Pi (a,4)=\left\vert 
\begin{tabular}{llll}
$a_{0}$ & $a_{2}$ & $a_{4}$ & $a_{6}$ \\ 
$a_{1}$ & $a_{3}$ & $a_{5}$ & $a_{7}$ \\ 
$a_{8}$ & $a_{10}$ & $a_{12}$ & $a_{14}$ \\ 
$a_{9}$ & $a_{11}$ & $a_{13}$ & $a_{15}$%
\end{tabular}%
\right\vert ,
\end{equation}%
which is equal to the determinant $N$ in \cite{Luque1}.

Now, suppose that $|\psi ^{\prime }\rangle $ and $|\psi \rangle $ are
equivalent under SLOCC. Then we get the following result: 
\begin{equation}
\Pi (a,n)=\Pi (b,n)\bigl[\det (\mathcal{A}_{1}) \cdots \det (\mathcal{A}_{n})%
\bigr]^{2^{(n-2)/2}},  \label{main-eq2}
\end{equation}%
where $\Pi (b,n)$ is obtained from $\Pi (a,n)$ by replacing $a$ by $b$. Eq. (%
\ref{main-eq2}) and $\Pi (a,n)$ are referred to as SLOCC equation and
polynomial of type II for even $n$ qubits, respectively. For $n=2$, Eq. (\ref%
{main-eq2}) can be verified by directly solving Eq. (\ref{neqcond}). For $%
n\geq 4$, we refer the reader to Appendix B for the proof.

It follows from Eq. (\ref{main-eq2}) that if one of $\Pi (a,n)$ and $\Pi
(b,n)$ vanishes while the other does not, then the state $|\psi ^{\prime
}\rangle $ is not equivalent to the state $|\psi \rangle $ under SLOCC.

Furthermore, it is trivial to see that $\Pi (a,n)$ vanishes for the $%
|GHZ\rangle $ and $|W\rangle $ states for $n>2$. For Dicke states $%
|l,n\rangle $ ($l\geq2$) for $n>2$, $\Pi (a,n)$ vanishes as well owing to
the fact that the second and third rows of $\Pi (a,n)$ are equal.

Consider the following two states 
\begin{eqnarray}
|\chi _{3}\rangle &=&(1/\sqrt{2^{n/2}})\biggl[%
\sum_{m=0}^{2^{n/2-1}-2}(|2^{n/2+1}m+4m\rangle +|2^{n/2+1}m+4m+3\rangle
)+|2^{n}-4\rangle -|2^{n}-1\rangle \biggr], \\
|\chi _{4}\rangle &=&(1/\sqrt{2^{n/2}})\biggl[%
\sum_{m=1}^{2^{n/2-1}-1}(|2^{n/2+1}m-4m+2\rangle +|2^{n/2+1}m-4m+1\rangle
)+|2^{n}-2^{n/2+1}+2\rangle  \notag \\
&-&|2^{n}-2^{n/2+1}+1\rangle \biggr].
\end{eqnarray}%
An argument analogous to the one in section 2 shows that $\Pi (a,n)$ does
not vanish for $|\chi _{3}\rangle $ or for $|\chi _{4}\rangle $. In light of
Eq. (\ref{main-eq2}), for $n>2$, the states $|\chi _{3}\rangle $ and $|\chi
_{4}\rangle $ are both different from the $|GHZ\rangle $, $|W\rangle $, and
Dicke states under SLOCC. It can be further demonstrated that $|\chi
_{3}\rangle $ and $|\chi _{4}\rangle $ are entangled, and that $|\chi
_{4}\rangle $ is equivalent to $|\chi _{3}\rangle $ under SLOCC. We
exemplify the result for the case of four qubits. We find that $|\chi
_{3}\rangle =(1/2)\bigl(|0\rangle +|3\rangle +|12\rangle -|15\rangle \bigr)$
and it was shown in \cite{LDF07b} that $|\chi _{3}\rangle $ is different
from the $|GHZ\rangle $, $|W\rangle $, and Dicke states under SLOCC.

\textsl{Remark 3.1}. In light of Eq. (\ref{main-eq1}), for $n>2$, $|\chi
_{3}\rangle $ is inequivalent to $|\chi _{1}\rangle $ under SLOCC, since we
can show that $\Theta(a,n)=0 $ for $|\chi _{3}\rangle $ and $\Theta
(a,n)\neq 0$ for $|\chi_{1}\rangle $.

\textsl{Remark 3.2.} For $|\chi _{3}\rangle $ SLOCC entanglement class,\ the
states $|\chi _{3}\rangle $ and $|\chi _{4}\rangle $ have the minimal number
of product terms (\emph{i.e.} $2^{n/2}$ product terms).

\section{SLOCC equation and polynomial of type III}

For the state $|\psi ^{\prime }\rangle $ of even $n$ qubits, let $\Gamma
(a,n)$ be the determinant of the coefficient matrix ( $2^{n/2}$ by $2^{n/2})$
which is partitioned into $2^{n/2+1}$ $1$ by $2^{n/2-1}$ blocks, \emph{i.e.} 
\begin{equation}
\Gamma (a,n)=\left\vert \left( 
\begin{tabular}{cc}
$\Gamma _{1}$ & $\Gamma _{1}^{\prime }$ \\ 
$\Gamma _{2}$ & $\Gamma _{2}^{\prime }$ \\ 
$\vdots $ & $\vdots $ \\ 
$\Gamma _{2^{n/2}}$ & $\Gamma _{2^{n/2}}^{\prime }$%
\end{tabular}%
\right) \right\vert ,
\end{equation}%
where the blocks $\Gamma _{i}$ and $\Gamma _{i}^{\prime }$, $%
i=1,2,\cdots,2^{n/2}$, satisfy that $\bigl\{\Gamma _{1}$ $\Gamma _{2} \cdots
\Gamma_{2^{n/2}}\bigr\}$ is just the coefficient vector $\bigl\{a_{0}$ $%
a_{1} \cdots a_{2^{n-1}-1}\}$, and $\{\Gamma _{1}^{\prime }$ $\Gamma
_{2}^{\prime } \cdots \Gamma _{2^{n/2}}^{\prime }\bigr\}$ is just the
coefficient vector $\bigl\{a_{2^{n-1}} $ $a_{2^{n-1}+1} \cdots a_{2^{n}-1}%
\bigr\}$.

To understand the structure of $\Gamma (a,n)$, we list $\Gamma (a,6)$ below: 
\begin{equation}
\left\vert 
\begin{tabular}{llllllll}
$a_{0}$ & $a_{1}$ & $a_{2}$ & $a_{3}$ & $a_{32}$ & $a_{33}$ & $a_{34}$ & $%
a_{35}$ \\ 
$a_{4}$ & $a_{5}$ & $a_{6}$ & $a_{7}$ & $a_{36}$ & $a_{37}$ & $a_{38}$ & $%
a_{39}$ \\ 
$a_{8}$ & $a_{9}$ & $a_{10}$ & $a_{11}$ & $a_{40}$ & $a_{41}$ & $a_{42}$ & $%
a_{43}$ \\ 
$a_{12}$ & $a_{13}$ & $a_{14}$ & $a_{15}$ & $a_{44}$ & $a_{45}$ & $a_{46}$ & 
$a_{47}$ \\ 
$a_{16}$ & $a_{17}$ & $a_{18}$ & $a_{19}$ & $a_{48}$ & $a_{49}$ & $a_{50}$ & 
$a_{51}$ \\ 
$a_{20}$ & $a_{21}$ & $a_{22}$ & $a_{23}$ & $a_{52}$ & $a_{53}$ & $a_{54}$ & 
$a_{55}$ \\ 
$a_{24}$ & $a_{25}$ & $a_{26}$ & $a_{27}$ & $a_{56}$ & $a_{57}$ & $a_{58}$ & 
$a_{59}$ \\ 
$a_{28}$ & $a_{29}$ & $a_{30}$ & $a_{31}$ & $a_{60}$ & $a_{61}$ & $a_{62}$ & 
$a_{63}$%
\end{tabular}%
\right\vert.
\end{equation}

Now, suppose that $|\psi ^{\prime }\rangle $ and $|\psi \rangle $ are
equivalent under SLOCC. Then we get the following result: 
\begin{equation}
\Gamma (a,n)=\Gamma (b,n)\bigl[\det (\mathcal{A}_{1})\cdots \det (\mathcal{A}%
_{n})\bigr]^{2^{(n-2)/2}},  \label{main-eq3}
\end{equation}%
where $\Gamma (b,n)$ is obtained from $\Gamma (a,n)$ by replacing $a$ by $b$%
. Eq. (\ref{main-eq3}) and $\Gamma (a,n)$ are referred to as SLOCC equation
and polynomial of type III for even $n$ qubits, respectively. For $n=2$, Eq.
(\ref{main-eq3}) can be verified by directly solving Eq. (\ref{neqcond}).
For $n\geq 4$, we refer the reader to Appendix C for the proof.

It follows from Eq. (\ref{main-eq3}) that if one of $\Gamma (a,n)$ and $%
\Gamma (b,n)$ vanishes while the other does not, then the state $|\psi
^{\prime}\rangle $ is not equivalent to the state $|\psi \rangle $ under
SLOCC.

Furthermore, it is trivial to see that $\Gamma (a,n)$ vanishes for the $%
|GHZ\rangle $ and $|W\rangle $ states for $n>2$. For Dicke states $%
|l,n\rangle $ ($l\geq 2$) for $n>2$, $\Gamma (a,n)$ vanishes as well owing to
the fact that the second and third columns of $\Gamma (a,n)$ are equal.

Consider the following two states 
\begin{eqnarray}
|\chi _{5}\rangle &=&(1/\sqrt{2^{n/2}})\biggl[%
\sum_{m=0}^{2^{n/2-1}-1}|(2^{n/2-1}+1)m\rangle
+\sum_{m=0}^{2^{n/2-1}-2}|(2^{n/2-1}+1)m+3\cdot 2^{n-2}\rangle
-|2^{n}-1\rangle \biggr], \\
|\chi _{6}\rangle &=&(1/\sqrt{2^{n/2}})\biggl[%
\sum_{m=1}^{2^{n/2-1}}|2^{n-1}+(2^{n/2-1}-1)m\rangle
+\sum_{m=1}^{2^{n/2-1}-1}|2^{n-2}+(2^{n/2-1}-1)m\rangle  \notag \\
&-&|2^{n-1}-2^{n/2-1}\rangle \biggr].
\end{eqnarray}%
An argument analogous to the one in section 2 shows that $\Gamma (a,n)$ does
not vanish for $|\chi _{5}\rangle $ or for $|\chi _{6}\rangle $. In light of
Eq. (\ref{main-eq3}), for $n>2$, $|\chi _{5}\rangle $ and $|\chi _{6}\rangle 
$ are both different from the $|GHZ\rangle $, $|W\rangle $, and Dicke states
under SLOCC. It can be further demonstrated that $|\chi _{5}\rangle $ and $%
|\chi _{6}\rangle $ are entangled, and that $|\chi _{6}\rangle $ is
equivalent to $|\chi _{5}\rangle $ under SLOCC. We exemplify the result for
the case of four qubits. We find that $\Gamma (a,4)=\Pi (a,4)$ and $|\chi
_{5}\rangle =|\chi _{3}\rangle $.

\textsl{Remark 4.1.} In light of Eqs. (\ref{main-eq1}) and (\ref{main-eq2}), 
$|\chi _{5}\rangle $ is inequivalent to $|\chi _{1}\rangle $ for $n>2$ or $%
|\chi _{3}\rangle $ for $n>4$ under SLOCC, since we can show that $\Theta
(a,n)=\Pi (a,n)=0 $ for $|\chi _{5}\rangle $, $\Theta (a,n)\neq 0$ for $%
|\chi _{1}\rangle $ and $\Pi (a,n)\neq 0$ for $|\chi _{3}\rangle $.

\textsl{Remark 4.2.} For $|\chi _{5}\rangle $ SLOCC entanglement class, the
states $|\chi _{5}\rangle $ and $|\chi _{6}\rangle $ have the minimal number
of product terms (\emph{i.e.} $2^{n/2}$ product terms).

\section{SLOCC equation and polynomial of type IV}

For the state $|\psi ^{\prime }\rangle $ of even $n$ qubits, let $\Omega
(a,n)$ be the determinant of the coefficient matrix ($2^{n/2}$ by $2^{n/2})$
which is partitioned into $2^{n/2+1}$ $1$ by $2^{n/2-1}$ blocks, \emph{i.e.} 
\begin{equation}
\Omega (a,n)=\left\vert \left( 
\begin{tabular}{cc}
$\Omega _{1}$ & $\Omega _{1}^{\prime }$ \\ 
$\Omega _{2}$ & $\Omega _{2}^{\prime }$ \\ 
$\vdots $ & $\vdots $ \\ 
$\Omega _{2^{n/2}}$ & $\Omega _{2^{n/2}}^{\prime }$%
\end{tabular}%
\right) \right\vert ,  \label{invariant-4}
\end{equation}%
where the blocks $\Omega _{i}$\ and $\Omega _{i}^{\prime }$ satisfy 
\begin{eqnarray}
\bigl\{\Omega_{1} \Omega_{2} \Omega_{5} \Omega_{6} \cdots \Omega_{4k+1}
\Omega_{4k+2}\cdots \Omega_{2^{n/2}-3} \Omega_{2^{n/2}-2}\bigr\} &=&\bigl\{%
a_{0},a_{2},\cdots a_{2^{n-1}-2}\bigr\}, \\
\bigl\{\Omega_{3} \Omega_{4} \Omega_{7} \Omega_{8}\cdots \Omega_{4k+3}
\Omega_{4k+4}\cdots \Omega_{2^{n/2}-1} \Omega_{2^{n/2}}\bigr\} &=&\bigl\{%
a_{1},a_{3},\cdots,a_{2^{n-1}-1}\bigr\}, \\
\bigl\{\Omega_{1}^{\prime } \Omega_{2}^{\prime } \Omega_{5}^{\prime }
\Omega_{6}^{\prime } \cdots \Omega_{4k+1}^{\prime } \Omega_{4k+2}^{\prime}
\cdots \Omega_{2^{n/2}-3}^{\prime } \Omega_{2^{n/2}-2}^{\prime}\bigr\} &=&%
\bigl\{a_{2^{n-1}},a_{2^{n-1}+2},\cdots,a_{2^{n}-2}\bigr\}, \\
\bigl\{\Omega_{3}^{\prime } \Omega_{4}^{\prime } \Omega_{7}^{\prime }
\Omega_{8}^{\prime } \cdots \Omega_{4k+3}^{\prime } \Omega_{4k+4}^{\prime }
\cdots \Omega_{2^{n/2}-1}^{\prime } \Omega_{2^{n/2}}^{\prime}\bigr\} &=&%
\bigl\{a_{2^{n-1}+1},a_{2^{n-1}+3},\cdots,a_{2^{n}-1}\bigr\},
\end{eqnarray}
for $0\leq k\leq 2^{n/2-2}-1$.

To understand the structure of $\Omega (a,n)$, we list $\Omega (a,6)$ below: 
\begin{equation}
\left\vert 
\begin{tabular}{llllllll}
$a_{0}$ & $a_{2}$ & $a_{4}$ & $a_{6}$ & $a_{32}$ & $a_{34}$ & $a_{36}$ & $%
a_{38}$ \\ 
$a_{8}$ & $a_{10}$ & $a_{12}$ & $a_{14}$ & $a_{40}$ & $a_{42}$ & $a_{44}$ & $%
a_{46}$ \\ 
$a_{1}$ & $a_{3}$ & $a_{5}$ & $a_{7}$ & $a_{33}$ & $a_{35}$ & $a_{37}$ & $%
a_{39}$ \\ 
$a_{9}$ & $a_{11}$ & $a_{13}$ & $a_{15}$ & $a_{41}$ & $a_{43}$ & $a_{45}$ & $%
a_{47}$ \\ 
$a_{16}$ & $a_{18}$ & $a_{20}$ & $a_{22}$ & $a_{48}$ & $a_{50}$ & $a_{52}$ & 
$a_{54}$ \\ 
$a_{24}$ & $a_{26}$ & $a_{28}$ & $a_{30}$ & $a_{56}$ & $a_{58}$ & $a_{60}$ & 
$a_{62}$ \\ 
$a_{17}$ & $a_{19}$ & $a_{21}$ & $a_{23}$ & $a_{49}$ & $a_{51}$ & $a_{53}$ & 
$a_{55}$ \\ 
$a_{25}$ & $a_{27}$ & $a_{29}$ & $a_{31}$ & $a_{57}$ & $a_{59}$ & $a_{61}$ & 
$a_{63}$%
\end{tabular}%
\right\vert.
\end{equation}

Now, suppose that $|\psi ^{\prime }\rangle $ and $|\psi \rangle $ are
equivalent under SLOCC. Then we get the following result: 
\begin{equation}
\Omega (a,n)=\Omega (b,n)\bigl[\det (\mathcal{A}_{1})\cdots\det (\mathcal{A}%
_{n})\bigr]^{2^{(n-2)/2}},  \label{main-eq4}
\end{equation}%
where $\Omega (b,n)$ is obtained from $\Omega (a,n)$ by replacing $a$ by $b$%
. Eq. (\ref{main-eq4}) and $\Omega (a,n)$ are referred to as SLOCC equation
and polynomial of type IV for even $n$ qubits, respectively. For $n=2$, Eq. (%
\ref{main-eq4}) can be verified by directly solving Eq. (\ref{neqcond}). For 
$n\geq 4$, we refer the reader to Appendix D for the proof.

It follows from Eq. (\ref{main-eq4}) that if one of $\Omega (a,n)$ and $%
\Omega (b,n)$ vanishes while the other does not, then the state $|\psi
^{\prime }\rangle $ is not equivalent to the state $|\psi \rangle $ under
SLOCC.

Furthermore, it is trivial to see that $\Omega (a,n)$ vanishes for the $%
|GHZ\rangle $ and $|W\rangle $ states for $n>2$. For Dicke states $%
|l,n\rangle $ ($l\geq 2$) for $n>2$, $\Omega (a,n)$ vanishes as well owing
to the fact that the second and third columns of $\Omega (a,n)$ are equal.

Consider the following state 
\begin{eqnarray}
|\chi _{7}\rangle &=&(1/\sqrt{2^{n/2}})\biggl[%
\sum_{m=0}^{2^{n/2-3}-1}(|2^{n/2+1}m+8m\rangle +|2^{n/2+1}m+8m+3\cdot
2^{n-2}\rangle )  \notag \\
&+&\sum_{m=0}^{2^{n/2-3}-1}(|(2m+1)2^{n/2}+8m+2\rangle
+|(2m+1)2^{n/2}+8m+2+3\cdot 2^{n-2}\rangle )  \notag \\
&+&\sum_{m=0}^{2^{n/2-3}-1}(|2^{n/2+1}m+8m+5\rangle +|2^{n/2+1}m+8m+5+3\cdot
2^{n-2}\rangle )  \notag \\
&+&\sum_{m=0}^{2^{n/2-3}-1}(|(2m+1)2^{n/2}+8m+7\rangle
+|(2m+1)2^{n/2}+8m+7+3\cdot 2^{n-2}\rangle )\biggr]  \notag \\
&-&(2/\sqrt{2^{n/2}})|2^{n}-1\rangle ,
\end{eqnarray}%
for $n\geq 6$ and $|\chi _{7}\rangle =(1/2)\bigl(|0\rangle +|6\rangle
+|9\rangle -|15\rangle \bigr)$ for $n=4$. An argument analogous to the one
in section 2 shows that $\Omega (a,n)$ does not vanish for $|\chi
_{7}\rangle $. In light of Eq. (\ref{main-eq4}), for $n>2$, $|\chi
_{7}\rangle $ is different from the $|GHZ\rangle $, $|W\rangle $, and Dicke
states under SLOCC. It can be further demonstrated that the state $|\chi
_{7}\rangle $ is entangled. In particular, for four qubits, it was shown in 
\cite{LDF07b} that $|\chi _{7}\rangle $ is different from the $|GHZ\rangle $%
, $|W\rangle $ and Dicke states under SLOCC. We further note that $|\chi
_{7}\rangle =$ $|\chi _{5}\rangle $ for the case of six qubits.

\textsl{Remark 5.1.} In light of Eqs. (\ref{main-eq1}), (\ref{main-eq2}),
and (\ref{main-eq3}), for $n>2$, $|\chi _{7}\rangle $ is inequivalent to $%
|\chi _{1}\rangle $, $|\chi _{3}\rangle $, or $|\chi _{5}\rangle $ ($n\neq 6 
$ for $|\chi _{5}\rangle $) under SLOCC, since we can show that $\Theta
(a,n)=\Pi (a,n)=\Gamma (a,n)=0$ for $|\chi _{7}\rangle $.

\textsl{Remark 5.2.} For $|\chi _{7}\rangle $ SLOCC entanglement class, the
state $|\chi _{7}\rangle $ has the minimal number of product terms (\emph{%
i.e.} $2^{n/2}$ product terms).

\section{Conclusion}

In this paper, for even $n$ qubits we have established four SLOCC equations
and constructed four SLOCC polynomials of degree $2^{n/2}$. For $n=4$, the
SLOCC polynomials of degree $2^{n/2}$ reduce to the polynomials of degree 4
in \cite{Luque1}. For $n\geq 6$, the four SLOCC polynomials are linearly
independent. The SLOCC equations can be exploited for SLOCC classification
of any even $n$ qubits. In light of the SLOCC equations, we have proposed
several different genuine SLOCC entanglement classes of even $n $ qubits and
showed that they are inequivalent to the $|GHZ\rangle $, $|W\rangle $, or $%
|l,n\rangle $ (the symmetric Dicke states with $l$ excitations) via the
vanishing or not of the polynomials.

The concurrence and three-tangle, which measure entanglement of two-qubit
and three-qubit states, have been known to be the absolute values of
hyperdeterminants for two and three qubits respectively \cite{Miyake}.
Recently, polynomial invariants have been proposed to construct entanglement
monotones. The absolute values of the polynomial invariants obtained in \cite%
{Luque1} may be used to construct entanglement measures of four-qubit
states. We expect that the absolute values of the polynomials in this paper
can be considered as entanglement measures.

\section*{Appendix A. The proof for SLOCC equation of type I}


\setcounter{equation}{0} \renewcommand{\theequation}{A\arabic{equation}}

\textsl{Proof}. We will prove Eq. (\ref{main-eq1}) by induction principle.
For the base case, letting $\mathcal{A}_{1}= \mathcal{A}_{2}=\cdots=\mathcal{%
A}_{n}=I$ in Eq. (\ref{neqcond}) yields $\Theta(a,n)=\Theta (b,n)$.

Let $|\phi \rangle =\sum_{i=0}^{2^{n}-1}c_{i}|i\rangle $ and 
\begin{equation}
|\phi \rangle =\underbrace{I\otimes \cdots \otimes I\otimes \mathcal{A}%
_{r+1}\otimes \cdots \otimes \mathcal{A}_{n}}_{n}|\psi \rangle .
\label{neq-2}
\end{equation}

Assume that $\Theta (c,n)=\Theta (b,n)\bigl[\det (\mathcal{A}_{r+1})\cdots
\det (\mathcal{A}_{n})\bigr]^{2^{(n-2)/2}}$, where $\Theta (c,n)$ is
obtained from $\Theta(a,n)$ by replacing $a$ by $c$. Next we will show that when 
\begin{equation}
|\psi ^{\prime }\rangle =\underbrace{I\otimes \cdots \otimes I\otimes 
\mathcal{A}_{r}\otimes \cdots \otimes \mathcal{A}_{n}}_{n}|\psi \rangle ,
\label{neq-3}
\end{equation}%
then 
\begin{equation}
\Theta (a,n)=\Theta (b,n)\bigl[\det (\mathcal{A}_{r})\cdots \det (\mathcal{A}%
_{n})\bigr]^{2^{(n-2)/2}}.
\end{equation}%
It is easy to see that $|\psi ^{\prime }\rangle $ $=\underbrace{I\otimes
\cdots \otimes I\otimes \mathcal{A}_{r}\otimes I\cdots \otimes I}_{n}|\phi
\rangle $. If we can prove that $\Theta (a,n)=\Theta (c,n)\bigl[\det (%
\mathcal{A}_{r})\bigr]^{2^{(n-2)/2}}$, then we can finish the induction.

For readability, let $\mathcal{A}_{l+1}=$ $\tau =\left( 
\begin{tabular}{ll}
$\tau _{1}$ & $\tau _{2}$ \\ 
$\tau _{3}$ & $\tau _{4}$%
\end{tabular}%
\right) $. Thus, we only need to prove that 
\begin{equation}
\Theta (a,n)=\Theta (c,n)\bigl[\det (\tau )\bigr]^{2^{(n-2)/2}},
\label{SLOCC-eq-1}
\end{equation}%
whenever $|\psi ^{\prime }\rangle $ and $|\phi \rangle $ satisfy the
following equation 
\begin{equation}
|\psi ^{\prime }\rangle =\underbrace{I\otimes \cdots\otimes I}_{l}\otimes
\tau \otimes \underbrace{I\otimes \cdots\otimes I}_{n-l-1}|\phi \rangle .
\label{induc-eq}
\end{equation}

From Eq. (\ref{induc-eq}), we obtain 
\begin{eqnarray}
a_{2^{n-l}k+s} &=&\tau _{1}c_{2^{n-l}k+s}+\tau _{2}c_{2^{n-l}k+2^{n-l-1}+s},
\label{coeff-0} \\
a_{2^{n-l}k+2^{n-l-1}+s} &=&\tau _{3}c_{2^{n-l}k+s}+\tau
_{4}c_{2^{n-l}k+2^{n-l-1}+s},  \label{coeff-1}
\end{eqnarray}%
where $0\leq k\leq 2^{l}-1$ and $0\leq s\leq 2^{n-l-1}-1$.

We distinguish two cases.

\textsl{Case 1.} $0\leq l\leq n/2-1$.

Let $A_{k,j}$ be a column of $\Theta (a,n)$ with entries $%
a_{2^{n-l}k+2^{n/2}j+q}$ where $0\leq q\leq (2^{n/2}-1)$, and let $%
A_{k,j}^{\ast }$ be a column obtained from $A_{k,j}$ by replacing each entry 
$a_{\eta }$ by $a_{\eta +2^{n-l-1}}$. Then, the columns of $\Theta (a,n)$
are (from left to right) 
\begin{equation}
\cdots, A_{k,j}, A_{k,j+1}, \cdots, A_{k,j}^{\ast }, A_{k,j+1}^{\ast},
\cdots, A_{k+1,j}, A_{k+1,j+1}, \cdots, A_{k+1,j}^{\ast },
A_{k+1,j+1}^{\ast}, \cdots,
\end{equation}
where $0\le k\le 2^l-1$ and $0\le j\le 2^{n/2-l-1}-1$.

Note that $2^{n/2}j+q\leq 2^{n-l-1}-1$. Substituting Eqs. (\ref{coeff-0})
and (\ref{coeff-1}) into $A_{k,j}$ and $A_{k,j}^{\ast }$ yields $%
A_{k,j}=\tau _{1}C_{k,j}+\tau _{2}C_{k,j}^{\ast }$ and $A_{k,j}^{\ast }=\tau
_{3}C_{k,j}+\tau _{4}C_{k,j}^{\ast }$, where $C_{k,j}$ and $C_{k,j}^{\ast }$
are obtained from $A_{k,j}$ and $A_{k,j}^{\ast }$ by replacing $a$ by $c$
respectively. We see that $C_{k,j}$ and $C_{k,j}^{\ast }$ are columns of $%
\Theta (c,n)$.

To compute $\Theta (a,n)$, we first let $\mathcal{T}_{k,j}$ be either $\tau
_{1}$ or $\tau _{2}$, and let $\mathcal{T}_{k,j}^{\ast }$ be either $\tau
_{3}$ or $\tau _{4}$. Let $U_{k,j}=C_{k,j}$ if $\mathcal{T}_{k,j}=\tau _{1}$%
, and $U_{k,j}=C_{k,j}^{\ast }$ otherwise. Further, let $U_{k,j}^{\ast
}=C_{k,j}$ if $\mathcal{T}_{k,j}^{\ast }=\tau _{3}$, and $U_{k,j}^{\ast
}=C_{k,j}^{\ast }$ otherwise. Due to the multilinear property of
determinant, $\Theta (a,n)$ is the sum of $2^{2^{n/2}}$ determinants, each
of which consists of columns (from left to right): 
\begin{eqnarray*}
& & \cdots,\mathcal{T}_{k,j}U_{k,j}, \mathcal{T}_{k,j+1}U_{k,j+1}, \cdots, 
\mathcal{T}_{k,j}^{\ast}U_{k,j}^{\ast }, \mathcal{T}_{k,j+1}^{%
\ast}U_{k,j+1}^{\ast }, \cdots, \mathcal{T}_{k+1,j}U_{k+1,j}, \mathcal{T}%
_{k+1,j+1}U_{k+1,j+1}, \\
& & \hskip 3 in \cdots, \mathcal{T}_{k+1,j}^{\ast}U_{k+1,j}^{\ast }, 
\mathcal{T}_{k+1,j+1}^{\ast}U_{k+1,j+1}^{\ast }, \cdots,
\end{eqnarray*}%
where $0\leq k\leq 2^{l}-1$ and $0\leq j\leq 2^{n/2-l-1}-1$.

Denote $t$ the product 
\begin{equation*}
\cdots \mathcal{T}_{k,j}\mathcal{T}_{k,j+1}\cdots \mathcal{T}_{k,j}^{\ast }%
\mathcal{T}_{k,j+1}^{\ast }\cdots \mathcal{T}_{k+1,j}\mathcal{T}%
_{k+1,j+1}\cdots \mathcal{T}_{k+1,j}^{\ast }\mathcal{T}_{k+1,j+1}^{\ast
}\cdots .
\end{equation*}%
Clearly, each of the $2^{2^{n/2}}$ determinants can be written in the form $%
t\cdot \Delta $. Associated with each $t$ is a determinant $\Delta $ which
consists of columns (from left to right): 
\begin{equation*}
\cdots ,U_{k,j},U_{k,j+1},\cdots ,U_{k,j}^{\ast },U_{k,j+1}^{\ast },\cdots
,U_{k+1,j},U_{k+1,j+1},\cdots ,U_{k+1,j}^{\ast },U_{k+1,j+1}^{\ast },\cdots ,
\end{equation*}%
where $0\leq k\leq 2^{l}-1$ and $0\leq j\leq 2^{n/2-l-1}-1$. 

We illustrate with an example. Let $t=\cdots\tau_{1}\tau_{1}\cdots\tau_{4}
\tau_{4}\cdots \tau_{1}\tau_{1}\cdots\tau_{4}\tau_{4}\cdots$, whose power
form is $(\tau_{1}\tau_{4})^{2^{(n-2)/2}}$, then $\Delta=\Theta (c,n)$.

For Eq. (\ref{SLOCC-eq-1}) to hold, we need the following 3 results.

\textsl{Result 1. }Given $t$ such that for some $k,j$, $\mathcal{T}%
_{k,j}=\tau _{1}$ and $\mathcal{T}_{k,j}^{\ast }=\tau _{3}$, or $\mathcal{T}%
_{k,j}=\tau_{2}$ and $\mathcal{T}_{k,j}^{\ast }=\tau _{4}$, then $\Delta $
vanishes.

\textsl{Proof}. If $\mathcal{T}_{k,j}=\tau _{1}$ and $\mathcal{T}%
_{k,j}^{\ast }=\tau _{3}$, then by definition $U_{k,j}=U_{k,j}^{\ast
}=C_{k,j}$. We immediately see that $\Delta $ vanishes since $\Delta $ has
two equal columns. Likewise, if $\mathcal{T}_{k,j}=\tau _{2}$ and $\mathcal{T%
}_{k,j}^{\ast }=\tau _{4}$, then by definition $U_{k,j}=U_{k,j}^{\ast
}=C_{k,j}^{\ast }$ and therefore $\Delta $ vanishes.

\textsl{Result 2. }Given $t$ such that for $0\leq k\leq 2^{l}-1$ and $0\leq j\leq 2^{n/2-l-1}-1$, 
$\mathcal{T}_{k,j}=\tau _{1}$ and $%
\mathcal{T}_{k,j}^{\ast }=\tau _{4}$, or $\mathcal{T}_{k,j}=\tau _{2}$ and $%
\mathcal{T}_{k,j}^{\ast }=\tau _{3}$ with $m$ occurrences for each of $\tau
_{2}$ and $\tau _{3}$, then $\Delta =(-1)^{m}\Theta (c,n)$.

\textsl{Proof}. If $\mathcal{T}_{k,j}=\tau _{1}$ and $\mathcal{T}%
_{k,j}^{\ast }=\tau _{4}$ for some $k,j$, then by definition $%
U_{k,j}=C_{k,j} $ and $U_{k,j}^{\ast }=C_{k,j}^{\ast }$. 
These two columns of $\Delta $ are already in order
and nothing needs to be done here. If $\mathcal{T}_{k,j}=\tau _{2}$ and $%
\mathcal{T}_{k,j}^{\ast }=\tau _{3}$, then by definition $%
U_{k,j}=C_{k,j}^{\ast }$ and $U_{k,j}^{\ast }=C_{k,j}$. To obtain $\Theta
(c,n)$ from $\Delta $, we need to interchange these two columns. It turns
out that $\Theta (c,n)$ can be obtained from $\Delta $ by interchanging two
coulmns for $m$ times, \emph{i.e.} $\Delta =(-1)^{m}\Theta (c,n)$.

\textsl{Result 3. }The number of $t$ such that its power form is $(\tau
_{1}\tau _{4})^{i}(\tau _{2}\tau _{3})^{2^{(n-2)/2}-i}$ is given by $\left(%
\begin{tabular}{c}
$2^{(n-2)/2}$ \\ 
$i$%
\end{tabular}%
\right)$.

\textsl{Proof}. With the help of \textsl{Result 1}, we only need to consider
those $t$ in which $\mathcal{T}_{k,j}=\tau _{1}$ and $\mathcal{T}%
_{k,j}^{\ast }=\tau _{4}$, or $\mathcal{T}_{k,j}=\tau _{2}$ and $\mathcal{T}%
_{k,j}^{\ast }=\tau _{3}$. In fact, we only need to count the number of
occurrences of $\tau _{1}$ and $\tau _{2}$ in $\mathcal{T}_{k,0}\cdots 
\mathcal{T}_{k,j}\cdots \mathcal{T}_{k,2^{n/2-l-1}-1}$, where $0\leq k\leq
2^{l}-1$ and $0\leq j\leq 2^{n/2-l-1}-1$. It is readily seen that there are $%
\left( 
\begin{tabular}{c}
$2^{(n-2)/2}$ \\ 
$i$%
\end{tabular}%
\right) $ such cases, each of which contains $i$ occurrences of $\tau _{1}$
and $(2^{(n-2)/2}-i)$ occurrences of $\tau _{2}$.

It follows immediately from \textsl{Result 2} and \textsl{Result 3} that the
sum of the $2^{2^{n/2}}$ determinants is given by $\Theta (c,n)\bigl[\det (%
\mathcal{\tau })\bigr]^{2^{(n-2)/2}}$. Therefore Eq. (\ref{SLOCC-eq-1})
holds.

\textsl{Case 2.} $n/2\leq l\leq (n-1)$.

Results analogous to \textsl{Result 1}, \textsl{Result 2} and \textsl{Result
3} can be derived by replacing \textquotedblleft column\textquotedblright\
by \textquotedblleft row\textquotedblright .

Combining the above two cases, Eq. (\ref{SLOCC-eq-1}) holds, and the proof
is complete.

\section*{Appendix B. The proof for SLOCC equation of type II}

\setcounter{equation}{0} \renewcommand{\theequation}{B\arabic{equation}}

\textsl{Proof}. By induction principle and the argument in Appendix A, we
only need to prove that $\Pi (a,n)=\Pi (c,n)\bigl[\det (\mathcal{\tau })%
\bigr]^{2^{(n-2)/2}}$ when $|\psi ^{\prime }\rangle $ and $|\phi \rangle $
satisfy Eq. (\ref{induc-eq}).

We distinguish three cases.

\textsl{Case 1.} $0\leq l\leq n/2-2$.

The proof is analogous to that in \textsl{case 2} in Appendix A by
investigating the rows of $\Pi (a,n)$.

\textsl{Case 2.} $n/2-1\leq l\leq n-2$.

The proof is analogous to that in \textsl{case 1} in Appendix A by
investigating the columns of $\Pi (a,n)$.

\textsl{Case 3.} $l=n-1$.

In this case, Eqs. (\ref{coeff-0}) and (\ref{coeff-1}) become 
\begin{eqnarray}
a_{2k}&=&\tau _{1}c_{2k}+\tau _{2}c_{2k+1},  \label{coeff-2} \\
a_{2k+1}&=&\tau _{3}c_{2k}+\tau _{4}c_{2k+1},  \label{coeff-3}
\end{eqnarray}%
where $0\leq k\leq 2^{n-1}-1$. The $(2r)th$ row of $\Pi (a,n)\ $is given by $%
\bigl(a_{2^{n/2}2r}$, $a_{2^{n/2}2r+2}, \cdots , a_{2^{n/2}(2r+2)-2}\bigr)$,
and the $(2r+1)th$ row of $\Pi (a,n)$ can be obtained from the $(2r)th$ row
by replacing each entry $a_{\eta }$ by $a_{\eta +1}$. The rest of the proof
is analogous to that in \textsl{case 2} in Appendix B.

\section*{Appendix C. The proof for SLOCC equation of type III}

\setcounter{equation}{0} \renewcommand{\theequation}{C\arabic{equation}}

\textsl{Proof}. By induction principle and the argument in Appendix A, we
only need to prove that $\Gamma (a,n)=\Gamma (c,n)\bigl[\det (\mathcal{\tau }%
)\bigr]^{2^{(n-2)/2}}$ when $|\psi ^{\prime }\rangle $ and $|\phi \rangle $
satisfy Eq. (\ref{induc-eq}).

We distinguish three cases.

\textsl{Case 1.} $l=0$.

The proof is analogous to that in \textsl{case 1} in Appendix A by
investigating the columns of $\Gamma (a,n)$.

\textsl{Case 2.} $1\leq l\leq n/2$.

The proof is analogous to that in \textsl{case 2} in Appendix A by
investigating the rows of $\Gamma (a,n)$.

\textsl{Case 3.} $n/2+1\leq l\leq n-1$.

The proof is analogous to that in \textsl{case 1} in Appendix A by
investigating the columns of $\Gamma (a,n)$.

\section*{Appendix D. The proof for SLOCC equation of type IV}

\setcounter{equation}{0} \renewcommand{\theequation}{D\arabic{equation}}

\textsl{Proof}. By induction principle and the argument in Appendix A, we
only need to prove that $\Omega (a,n)=\Omega (c,n)\bigl[\det (\mathcal{\tau }%
)\bigr]^{2^{(n-2)/2}}$ when $|\psi ^{\prime }\rangle $ and $|\phi \rangle $
satisfy Eq. (\ref{induc-eq}).

We distinguish four cases.

\textsl{Case 1.} $l=0$.

The proof is analogous to that in \textsl{case 1} in Appendix A by
investigating the columns of $\Omega (a,n)$.

\textsl{Case 2.} $1\leq l\leq n/2-1$.

The proof is analogous to that in \textsl{case 2} in Appendix A by
investigating the rows of $\Omega (a,n)$.

\textsl{Case 3.} $n/2\leq l\leq n-2$.

The proof is analogous to that in \textsl{case 1} in Appendix A by
investigating the columns of $\Omega (a,n)$.

\textsl{Case 4.} $l=n-1$.

In this case, Eqs. (\ref{coeff-0}) and (\ref{coeff-1}) become Eqs. (\ref%
{coeff-2}) and (\ref{coeff-3}). Consider the $(4k+1)th$, the $(4k+2)th$, the 
$(4k+3)th$, and the $(4k+4)th$ ($0\leq k\leq 2^{n/2-2}-1$) rows of $\Omega
(a,n)$, respectively. The rest of the proof is analogous to that in \textsl{%
case 2} in Appendix A.

\textbf{\ }

\end{document}